\documentclass{ws-ijgmmp}
\usepackage{notoccite}
\usepackage[usenames,dvipsnames,svgnames]{xcolor}
\usepackage[colorlinks=true,
      linkcolor=red,
      urlcolor=gray,
      citecolor=blue]{hyperref}
\usepackage[]{cite}
\usepackage{float}

\usepackage[colorinlistoftodos,prependcaption,textsize=tiny]{todonotes}
\newcounter{mycomment}

\begin{document}

\markboth{Dario Corona, Roberto Giamb\`o, Orlando Luongo}
{Entanglement in model independent cosmological scenario}

\catchline{}{}{}{}{}

\title{Motion of test particles in quasi anti-de Sitter regular black holes  }

\author{Dario Corona}
\address{School of Science and Technology, University of Camerino, Via Madonna delle Carceri, Camerino, 62032, Italy.}

\author{Roberto Giamb\`o}
\address{School of Science and Technology, University of Camerino, Via Madonna delle Carceri, Camerino, 62032, Italy.\\
Istituto Nazionale di Fisica Nucleare, INFN, Sezione di Perugia, 06123 Perugia, Italy.\\
IAPS, INAF - Tor Vergata, 00133 Roma, Italy.}

\author{Orlando Luongo}
\address{School of Science and Technology, University of Camerino, Via Madonna delle Carceri, Camerino, 62032, Italy.\\
Istituto Nazionale di Fisica Nucleare, INFN, Sezione di Perugia, 06123 Perugia, Italy.\\
SUNY Polytechnic Institute, 13502 Utica, New York, USA.\\
INAF - Osservatorio Astronomico di Brera, 20121 Milano, Italy.\\
Al-Farabi University, Al-Farabi, 71, 050040 Almaty, Kazakhstan.}


\maketitle

\begin{abstract}
We explore the characteristics of two novel regular spacetimes that exhibit a non-zero vacuum energy term, under the form of a (quasi) anti-de Sitter phase.
Specifically, the first metric is spherical, while the second, derived by applying the generalized Newman-Janis algorithm to the first, is axisymmetric. 
We show that the equations of state of the effective fluids associated with the two metrics  asymptotically tend to negative values, resembling quintessence.
In addition, we study test particle motions, illustrating  the main discrepancies among our models and more conventional metrics exhibiting non-vanishing anti-de Sitter phase.
\end{abstract}



\section{Introduction}

{ Incorporating the cosmological constant, $\Lambda$, or more broadly, the dark energy term into spacetimes, is currently the subject of intense research \cite{UN}. Indeed, characterizing how $\Lambda$ influences the large-scale expansion is a crucial challenge of modern cosmology and, so, understanding how dark energy  behaves in proximity of compact objects, i.e., around spacetime solutions, would shed light on the nature of dark constituents\footnote{Specifically, in the cosmic enigma, $\Lambda$ is a constant term reflecting vacuum energy, originating from quantum fluctuations \cite{Martin:2012bt}. According to this, the cosmological constant signifies a manifestation of dark energy that remains invariant over time, while dark energy is conventionally modeled using barotropic fluids, showing the cosmic acceleration at a specific transition time \cite{Capozziello:2021xjw,Farooq:2013hq,Capozziello:2014zda,Farooq:2016zwm,Muccino:2022rnd,Alfano:2023evg}.
} \cite{dpa1}. Alternatively, studying the nature of solutions whose fluids behaves in some regimes as exotic dark fluids would open new avenues in the realm of spacetime solutions characterizing compact objects \cite{dpa2,dpa3}. }

{ Among all the possible approaches to the dark sector, an intriguing possibility is offered by scalar fields that act as effective fluids, whose equation of state is dominant over barotropic fluids, see e.g. Refs. \cite{Linder:2008ya,Giambo:2023iuy,Capozziello:2019cav}.
The simplest scalar field used in cosmology is \emph{quintessence}, characterized by a positive kinetic term coupled with a potential that remains unspecified.
The effects of this potential, in a slow roll regime, mimes the $\Lambda$ equation of state and, so, it appears quite interesting to find spacetime solutions that asymptotically reproduce such effects, providing non-zero de Sitter phases.}

In this respect, the study of black holes has gained significant importance in light of modern discoveries such as gravitational waves and black hole shadows \cite{Abbott2016,EventHorizonTelescope:2019dse}. Consequently, as above reported, examining the behavior of black holes within quintessence fields or, more broadly, within dark energy environments, can offer valuable insights into the \emph{interaction between black holes and dark energy} \cite{Zhang:2023neo}.

To this end, initial attempts to incorporate black holes into quintessence environments were made \cite{Kiselev:2002dx}, but they resulted in dark energy fluids exhibiting distinct radial and tangential pressures \cite{Visser:2019brz}. This contrasts with the cosmological principle that assumes identical radial and tangential pressures as functions of cosmic time only.

Furthermore, recent developments have unveiled the possibility of non-singular black hole configurations, predicting the theoretical existence of \emph{regular black holes} \cite{Lan:2023cvz,Torres:2022twv,Malafarina:2022wmx,unoinpiu}. These solutions, derived from Einstein's field equations, introduce notable enhancements to the properties of black holes \cite{Luongo:2023aib,Boshkayev:2023fft,Luongo:2023jyz}. One intriguing characteristic of regular black holes is their ability to \emph{carry vacuum energy}. For example, this property has been extensively explored through the study of the \emph{Hayward solution} \cite{Hayward:2005gi} and its extensions \cite{Mosani:2023awd,Bambi:2013ufa}. 

{ Regular black holes can have origins in non-linear electrodynamics \cite{tres1,tres2} and their extensive use in describing the physics of compact objects has been recently debated, emphasizing how general relativity can naturally predict them \cite{tres4,tres5}. }

{ Motivated by the above considerations, we here focus on two particular spacetimes, exhibiting a nonzero term of vacuum energy, under the form of a (quasi) anti-de Sitter phase. The first solution is spherical and derives from considering a $00$ term that is purely de Sitter, solving the Einstein equations without incorporating the Schwarzschild coordinates. The second metric, written in Boyer--Linquist coordinates, introduces a further quadrupole term, through the application of a generalized version of the Newman-Janis algorithm  \cite{aaa,Azreg-Ainou:2014aqa}, that we hereafter refer to as \textit{generalized Newman--Janis} (GNJ) algorithm. The corresponding asymptotic properties of our spacetimes are thus studied, remarking under which conditions the equations of state of the effective fluids, associated with our metrics, resemble quintessence at very large radii.  Further, we focus on the forms of horizons and geodesics, exploring the motion of test particles within the two metrics. Our findings indicate that these two solutions are novel and appear regular  with a (quasi) anti-de Sitter phase, somehow \emph{transporting} a vacuum energy-like contribution that resembles quintessence asymptotically.

The paper is structured as follows. In Sect. \ref{sezione2}, we review our spherical spacetime warm up, emphasizing how to construct the axisymmetric spacetime through the GNJ algorithm. In Sects. \ref{sezione3} and \ref{sezione4}, we  evaluate some particular features related to the two metrics under exam, whereas in Sect. \ref{sezione5}, we investigate the test particle motion for the metrics introduced before.  Finally, in Sect. \ref{sezione7}, we develop conclusions and perspectives of our work. }

\section{Geometric warm up}\label{sezione2}

Using the coordinate system, $(t,r,\theta,\varphi)$, the perfect fluid stress-energy
tensor can be recast as
\begin{equation}
\left[ T_{\nu}^{\mu} \right] =
\left[\begin{array}{cccc}
-\rho &       &      &     \\
      & P_{r} &      &     \\
      &       & P_{t} &     \\
      &       &      & P_{t}
\end{array}
\right] \,\,,
\label{emt}
\end{equation}
where the radial pressure, $P_r$, is different, in general, from the tangential pressure, $P_t$.

Given a static spherically symmetric solution,
\begin{equation}\label{3}
    d s^2 = -f(r)dt^2 + \frac{d r^2}{g(r)} + h(r)(d \theta^2+\sin^2\theta d \phi^2),
\end{equation}
one can assume $h(r)=r^2$ to guarantee that the volume is exactly that of a three-dimensional sphere. 

Moreover, from Eq. \eqref{3}, we can easily generate a stationary rotating solution using the above $h(r)$ \cite{aaa}. To do so, we obtain the metric, 
\begin{align}\label{4}
    ds^2 = &-\frac{f (g r^2+a^2 \cos^2\theta ) \Psi }{(\sqrt{g} r^2+a^2 \sqrt{f} \cos^2\theta )^2}\,
    dt^2
    +\frac{\Psi }{g r^2+a^2}\,dr^2
    \nonumber\\
    &-2 a \sin ^2\theta\Big[\frac{\sqrt{g} \sqrt{f} r^2-g\,f r^2}{(\sqrt{g} r^2+a^2 \sqrt{f} \cos ^2\theta )^2}\Big]\Psi dt d\phi
    +\Psi d \theta^2
    \nonumber\\
    &+\Psi\sin ^2\theta \Big\{1+a^2 \sin ^2\theta\Big[\frac{2 \sqrt{g} \sqrt{f} r^2-g\,f r^2
    +a^2 f \cos ^2\theta }{(\sqrt{g} r^2+a^2 \sqrt{f}\cos ^2\theta )^2}\Big]\Big\} d \phi^2,
\end{align}
{ where  $\Psi(r,\theta)$ is a function to fix, depending on the further rotating parameter, $a$, that can be fixed by solving the field equations. 

As a natural example, if $f(r)=g(r)$, a possibility is offered by $\Psi(r,\theta)=r^2+a^2 \cos^2\theta$.
More precisely, to obtain a black hole that mimes a dark energy fluid we might require:
\begin{itemize}
    \item[-] Isotropy of pressures, i.e., in fulfillment of the cosmological principle, the corresponding fluid appears physically motivated only if the tangential and radial pressures are the same. 
    \item[-] The density, pressure and equation of state might be independent from the radial coordinate. Again, the cosmological principle does not permit, in fact, to the component of the energy-momentum tensor to depend on $r$. To address this issue, since by construction the quantities might depend on $r$, one can focus on asymptotic regimes only, investigating whether the thermodynamic quantities are still coordinate-dependent or not, at very large radii. 
    \item[-] The equation of state, $w\equiv P/\rho$, might violate the Zeldovich limit to reproduce dark energy, namely $w\in[-1,0]$, whereas in the case $w=-1$ the fluid reproduces quintessence. 
    \item[-] The sound speed might be positive definite in order to fulfill stability in perturbation theory, i.e., to guarantee that no instabilities may occur as due to the dark energy fluid. 
\end{itemize}

Clearly, the energy-momentum tensor does not provide a \emph{physical} dark energy fluid as commonly used in cosmology, but rather a fluid that resembles the same properties of dark energy/quintessence.

We are thus interested in finding possible solutions to apply to compact objects by virtue of the above properties, by determining the energy-momentum tensor components, in both  the spherical and axisymmetric cases.  
}

\section{Spherical symmetry with asymptotic quintessence}\label{sezione3}

{ The simplest case corresponds to a spherically-symmetric solution. Thus, writing the Einstein equations,}
\begin{equation}
G_\nu^\mu\equiv R_{\nu}^{\mu} - \frac{1}{2} R \, \delta_{\nu}^{\mu} 
= 8\pi T_{\nu}^{\mu} \,\,,
\label{einstein_equations}
\end{equation}
with $G_\nu^\mu$ the Einstein tensor and $\delta^\mu_\nu$ the Kronecker delta, the simplest regular solution that transports vacuum energy is the de Sitter spacetime, where
\begin{equation}
f=g=1-\frac{\Lambda}{3} r^2\,.
\end{equation}

A further step consists in postulating 
\begin{equation}\label{eh}
    f(r)=1-\frac{\Lambda}{3}r^2\neq g(r),
\end{equation}
and to solve Eqs. \eqref{einstein_equations} to find $g(r)$.

Utilizing this recipe, we evaluate the Einstein equations based on Eq. \eqref{eh},
\begin{subequations}
\begin{align}
&{T_t}^t =\frac{(r g(r))^\prime -1  }{2r^2}\,,\\[2mm]
&{T_r}^r = \frac{3- \Lambda r^{2}  - 3(1 - 
    \Lambda r^{2 }  ) g(r)}{r^2 ( \Lambda r^{2} -3)},    \\[2mm]
&{T_\theta}^\theta = \frac{    
    (3 - r^{2} \Lambda) (3 - 
    2r^{2} \Lambda) g^\prime(r)-2r  \Lambda (6 -     r\Lambda) g(r)}{2r ( \Lambda r^{2} -3)^2}\,, 
\end{align}
\end{subequations}
with ${T_\phi}^\phi  = {T_\theta}^\theta$ and the prime indicating derivative with respect to the radial coordinate, $r$.

{ As previously stated, in order to guarantee isotropy on pressure, we require $P_r=P_t=P$, giving rise to the solution firstly found in Ref. \cite{GLmetric},
\begin{eqnarray}\label{eq:g11-eps0}
g(r)=\frac{\left(1+k_0 r^2\right) \left(1-\tfrac\Lambda 3  r^2\right)}{1-\tfrac23 \Lambda  r^2}\,.
\end{eqnarray}
Remarkably, requiring $g=1$ at $r=0$ and $g\sim r^2$, at $r\rightarrow\infty$, the natural recipe is to take $k_0\neq0$.}

\subsection{Comparing our solution with anti-de Sitter spacetime}

{ The anti-de Sitter spacetime is particularly simply in framing the corresponding energy-momentum tensor, 
\begin{equation}
T^{\mu}_\nu=-\Lambda\delta^\mu_\nu\,.
\end{equation}
while our solution produces the following outcomes,
\begin{subequations}    
\begin{align}
\rho &= \frac{\Lambda  \left(2 \Lambda  r^2-9\right)-3 k_0 \left(2 \Lambda ^2 r^4-7 \Lambda  r^2+9\right)}{\left(3-2 \Lambda  r^2\right)^2}, 
\label{eq:rho-eps0}\qquad\qquad\quad\,\,\,\\ 
P &= \frac{3 k_0 \left(\Lambda  r^2-1\right)+\Lambda }{2 \Lambda  r^2-3}\,.\label{eq:P-eps0}
\end{align}
\end{subequations}

At infinity, the behaviors of these two solutions appear interesting. While the density and pressure for the de Sitter spacetime are always constant, our spacetime reproduces stable $\rho$ and $P$ at large radii, with the characteristic of being independent on the radial coordinate at large radii. Thus, the corresponding equation of state resembles that of a cosmological constant asymptotically and, in this respect, one can conclude that the solution transports a de Sitter phase, namely contains a vacuum energy term, that dominates at $r\rightarrow\infty$.

Indeed, a \emph{genuine de Sitter} solution is recovered from Eq. \eqref{eq:g11-eps0} if
\begin{equation}\label{eq:k-dS}
g\sim \frac{k_0\,r^2}{2}\rightarrow\,k_0\simeq -\frac23\Lambda\,,
\end{equation}
while  around $r=0$, up to the second order in $r$, we have
\begin{equation}\label{gasint}
g\sim 1+\left({k_0} + \frac{\Lambda}{3}\right) r^2+\ldots\,.
\end{equation}
Moreover,  the Ricci scalar reads
\begin{equation}\label{ricci}
R=-
\frac{6 k_0 \left(4 \Lambda ^2 r^4-11 \Lambda  r^2+9\right)+4 \Lambda ^2 r^2}{\left(3-2 \Lambda  r^2\right)^2},
\end{equation}
while for the de Sitter solution appears simpler, 
\begin{equation}\label{ricciSOL}
R^{dS}=4\Lambda,
\end{equation}
implying that 
\begin{equation}
\lim_{r\rightarrow\infty}\left(R-R^{dS}\right)=-6 k_0 - 4 \Lambda,
\end{equation}
emphasizing the intriguing solution, $R=R_{dS}$, as 
\begin{equation}
k_0=-\frac{2}{3}\Lambda,
\end{equation}
valid, in particular, at all radii, i.e., not only asymptotically, as confirmed in Eq. \eqref{eq:k-dS}.

Clearly, we are forced to stress that both $1-{\tfrac\Lambda 3} r^2>0$ and $g(r)$ should have the same sign.
Thus, we obtain
$\frac{ 1+k_0 r^2}{3-2 \Lambda  r^2}>0$ and, moreover, to require a regular black hole solution, our spacetime might be anti-de Sitter, namely $\Lambda<0$. 
}

\section{Introducing vacuum energy for the axisymmetric case}\label{sezione4}

In the axisymmetric case, a slight complexity reduction occurs if coordinates $(t,r,y,\phi)$ are considered, where 
$y = \cos\theta$. 

Thus, we keep on denoting by $\Psi$ the free function introduced in Eq. \eqref{4}, that now reads $\Psi=\Psi(r,y)$,  containing the rotation parameter, $a$.

{ In accordance with the metric discussed in the previous section, we may recast
\begin{equation}
    g(r) = f(r) \frac{1 + k_0r^2}{1-\tfrac23\Lambda r^2},
\end{equation}
where $f(r)$ is given in Eq. \eqref{eh}.
Consequently, the metric in Eq. \eqref{4} becomes
\begin{multline}\label{eq:GJN}
ds^2=\Psi(r,y)\left\{ 
    -\frac{\left(\frac{\left(k r^4+r^2\right) \left(\Lambda  r^2-3\right)}{2 \Lambda  r^2-3}+\alpha ^2 y^2\right)}{\left(r^2 \sqrt{\frac{3 k r^2+3}{3-2 \Lambda  r^2}}+\alpha ^2 y^2\right)^2}\mathrm dt^2 \right.\\
    + \frac{1}{\alpha ^2+\frac{\left(k r^4+r^2\right) \left(\Lambda  r^2-3\right)}{2 \Lambda  r^2-3}}\,\mathrm dr^2
    +\frac{1}{1-y^2}\mathrm dy^2\\
+\left(1-y^2\right) \left(1+\frac{\alpha ^2 \left(1-y^2\right) \left(2 r^2 \sqrt{\frac{3 k r^2+3}{3-2 \Lambda  r^2}}-\frac{\left(k r^4+r^2\right) \left(\Lambda  r^2-3\right)}{2 \Lambda  r^2-3}+\alpha ^2 y^2\right)}{\left(r^2 \sqrt{\frac{3 k r^2+3}{3-2 \Lambda  r^2}}+\alpha ^2 y^2\right)^2}\right)\,\mathrm d\phi^2\\
\left.-\frac{2 \alpha  r^2 \left(1-y^2\right)   \left(\sqrt{\frac{3 k r^2+3}{3-2 \Lambda  r^2}}-\frac{\left(k r^2+1\right) \left(\Lambda  r^2-3\right)}{2 \Lambda  r^2-3}\right)}{\left(r^2 \sqrt{\frac{3 k r^2+3}{3-2 \Lambda  r^2}}+\alpha ^2 y^2\right)^2}\,\mathrm dt \mathrm d\varphi\right\}.
\end{multline}

The conformal factor $\Psi(r,y)$ in Eq. \eqref{eq:GJN} can be determined demanding suitable hypotheses on the energy-momentum tensor of the solution. 
A noticeable example is when the source term is postulated as a rotating fluid around the $z$ axis. This can be obtained as follows: once the position $\Delta=g(r)r^2+\alpha^2$ is made, and  the orthonormal basis $\{e_\alpha\}$
\begin{subequations}
    \begin{align}
e_t&=\frac1{\sqrt{\Psi\Delta}}\left((r^2\sqrt{g/f}+\alpha^2)\partial_t+a\,\partial_\phi\right), \qquad e_r=\sqrt\frac{\Delta}{\Psi}\,\partial_r,\\
e_y&=\frac{1}{\sqrt{\Psi}}\,\partial_y,\qquad e_\phi =-\frac{1}{\sqrt\Psi}\left(\alpha\sqrt{1-y^2}\,\partial_t+\frac{1}{\sqrt{1-y^2}}\partial_\phi\right),
\end{align}
\end{subequations}
is introduced, see Ref. \cite[Eq. (16)]{Azreg-Ainou:2014aqa}, one demands that the source energy-momentum tensor is diagonal with respect to this basis, see \cite[Eq. (17)]{Azreg-Ainou:2014aqa}. 
It can be shown that a necessary and sufficient condition to this requirement is that $G_{ry}=0$ and $G^t_\phi=(y^2-1)(r^2\sqrt{g/f}+\alpha^2) G^\phi_t$, that in turn induces two different conditions on $\Psi$, if the seed spherical metric is completely fixed. 

In Refs. \cite{aaa,Azreg-Ainou:2014aqa}, it is observed that when $g(r)=f(r)$ the choice
\begin{equation}
\Psi(r,y)=r^2+\alpha^2 y^2,    
\end{equation}
satisfies both the conditions at the same time. 
However, in the general case where $g(r) \ne f(r)$, as discussed in the previous section, this choice introduces shear stresses and momentum density terms into the energy--momentum tensor.
Indeed, writing $T$ as
\begin{equation}
    T={\widetilde T}^{\mu\nu}\,e_\mu\otimes e_\nu,
\end{equation}
besides the energy density $\rho={\widetilde T}^{00}$ and the pressures $p_i={\widetilde T}^{ii}$, $i=1,2,3$, also nontrivial  shear ${\widetilde T}^{12}={\widetilde T}^{21}$ and momentum density ${\widetilde T}^{30}={\widetilde T}^{03}$ will be present.
Noticeably enough, the above choice of $\Psi$ induces the following asymptotic behavior at infinity
\begin{equation}\label{eq:quint}
\lim_{r\to\infty}\frac{P_r+P_{t,1}+P_{t,2}}{3\,\rho}=-1,
\end{equation}
similarly to the spherical case, with the difference that we have here two distinct contributions to the tangential pressure, namely $P_{t,1}$ and $P_{t,2}$.
The full expressions of our pressure and density are reported in Appendix A.
}

\section{Geodesic test particle motion}\label{sezione5}

{ In the above sections, we focused on the mathematical structures of our solutions, emphasizing the thermodynamic features of our solutions.
It is now interesting to go further and investigate the motion of test particles.} 

\subsection{Spherical symmetry}

\begin{figure*}[h]
  \centering
    \includegraphics[width=.6\linewidth]{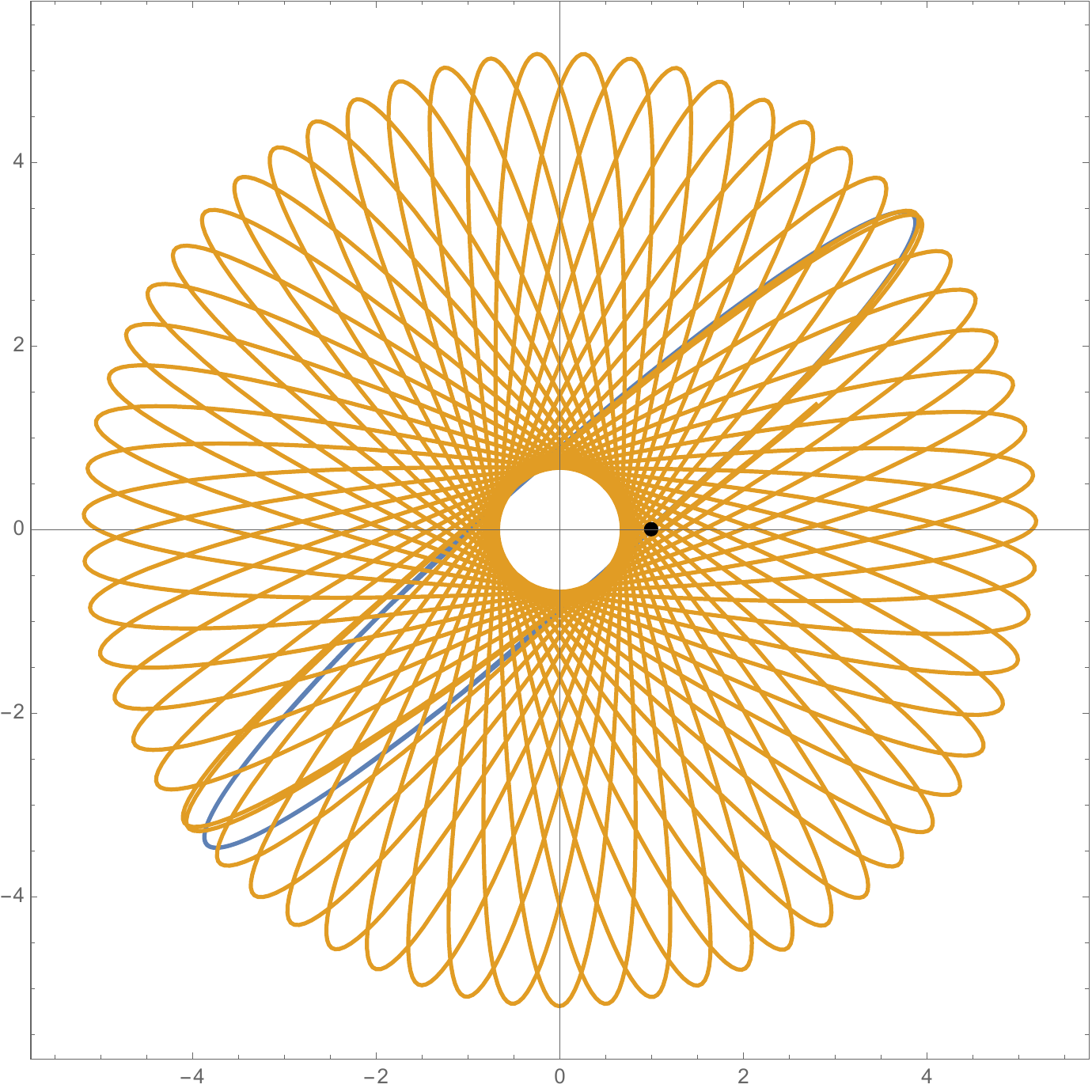}
  \hspace{0mm}\\
  \caption{Motion in the plane $\theta=\pi/2$ of  timelike geodesics with $E=4, L=2$. Here $\Lambda$ is set to -1 and $k=0.7$, not far away from $2/3$ that is the value yielding exactly the anti-de Sitter spacetime, see Eq. \eqref{eq:k-dS}. The blue closed curve is an anti-de Sitter timelike geodesic, and the orange one is a geodesic for the metric discussed above.}
  \label{geo_sph}
\end{figure*}

Let us first consider timelike geodesic in spherical symmetry. Parameterizing with the affine parameter, $\tau$, we have $(t(\tau), r(\tau), \theta(\tau), \phi(\tau))$ and, so, since we are dealing with a spherical configuration, as well known the existence of two (spacelike) Killing fields, namely $\sin \phi \partial_\theta,  + \cot \theta  \cos \phi \partial_\phi$ and $\cos \phi \partial_\theta - \cot \theta  \sin \phi \partial_\phi$ gives that, up to rotations, one can suppose that the geodesic lies on the equatorial plane $\theta=\pi/2$.
By means of this hypothesis, the geodesic equations read, in analogy to the spherical case,
\begin{align}
t^\prime(\tau)&=\frac{3E}{3 - r^2 \Lambda}\,,\\
\phi^\prime(\tau)&=\frac{L}{r^2}\,.
\end{align}
where primes indicate the derivatives with respect to the affine parameter, $\tau$. 
The last conditions suggest $\tau$ as proper time of the timelike geodesics, yielding the following relation to hold,
\begin{equation}
1-\frac{E^2}{f}+\frac{L^2}{r^2} + \frac{(r^\prime)^2}{g}=0\,.
\end{equation}
The admissible region, depending on the free parameters $ \Lambda$ and $k_0$, and, furthermore, on the constant of motion $E$ and $L$, with $r^2=x^2+y^2$, requires 
\begin{equation}
1 - \frac{E^2}{1 - {\tfrac\Lambda 3} (x^2 + y^2) + 
   L^2/(x^2 + y^2)} < 0.
\end{equation}
In other words, considering the effective potential 
\begin{equation}
V=\left(1 - \frac{\Lambda}{3} r^2\right) \left(1 + \frac{L^2}{r^2}\right),
\end{equation}
it might be $E^2\geq V$, if we require admissible solutions, similarly to the Schwarzschild case, with the difference that here we do not have any horizons. Indeed, the solution is regular everywhere, and then one may have geodesics arbitrarily close to the centre of symmetry $r=0$, choosing $E$ sufficiently large, as confirmed in Fig. \ref{geo_sph}.

As commonly known, we are primarily interested in the orbits $r(\phi)$. We denote $R(\phi)=r(\tau(\phi))$, then $r^\prime(\tau)=R^\prime(\phi)\phi^\prime(\tau)=\frac{R^\prime(\phi)L}{R (\phi)^2}$, therefore as in the Kepler problem we set $u(\phi)=(R(\phi))^{-1}$ and, then, $R'(\phi)=-\frac{u^\prime(\phi)}{u (\phi)^2}$, and differentiating with respect to $\phi$, we obtain the equation of timelike geodesic under the formal form, $u^{\prime\prime}(\phi)=W(\phi,u(\phi),u'(\phi))$, where the function $W(\phi,u,u^\prime)$ is given by

\begin{align}\label{WgeoSph}
W(\phi,u,u^\prime)=(k_0 + u(\phi)^2)&\left[\frac{\Lambda (3 E^2 + L^2 \Lambda) - 
  6 L^2 \Lambda u(\phi)^2 + 9 L^2 u(\phi)^4}{L^2 (2 \Lambda^2 u(\phi) -    9 \Lambda u(\phi)^3 + 9 u(\phi)^5)}\right]\\
-(\phi^{\prime})^2&\left[\frac{2 k_0 \Lambda^2 - 6 k_0 \Lambda u(\phi)^2 + 
   3 (3 k_0 + \Lambda) u(\phi)^4 }{u(\phi) (k_0 + u(\phi)^2) (2 \Lambda -    3 u(\phi)^2) (\Lambda - 3 u(\phi)^2)}\right],
\nonumber
\end{align}
where, interestingly, nonlinearities appear also in the first order derivative $u'(\phi)$, unlike the classical Schwarzschild spacetime. Accordingly, the study of geodesics can be performed by varying either the metric parameters or the constants of motion.

Additionally, the interval amplitude in which the solution exists might require peculiar attention, as well as initial conditions that, as due to the nonlinearities above cited, can significantly change the shape of curves derived from Eq. \eqref{WgeoSph}. In Fig. \ref{geo_sph}, we report an explicit representation of our solutions, showing the main departures with the timelike geodesics in the anti-de Sitter case -- as known, yet noticeably,  timelike geodesics in this latter case are \textit{closed} curves.  

\subsection{Axial symmetry}

For the sake of simplicity, we  focus on axial symmetry in the equatorial plane, where $\cos\theta=y=1$. This allows for a similar analysis as in the preceding section. However, a significant limitation arises from the reduced degrees of symmetry, rendering this assumption no more generic.

We set
$\dot\gamma(\tau) = \big(\dot{t}(\tau),\dot{r}(\tau), 0, \dot\phi(\tau)\big)$ and using $g(\dot\gamma,\partial_t) = - E$ and 
$g(\dot\gamma, \partial_\phi) = L$, 
we obtain
\begin{multline}
    \label{eq:equaGEO-tprime}
    \dot{t}
    = \frac{
    (3 - \Lambda r^2)(1+k_0r^2) - \sqrt{(3+3k_0r^2)(3-2r^2)}
    }
    {
    a^2 (3 - 2r^2) + (3-\Lambda r^2)(1+k_0r^2)r^2
    } aL \\
    + \frac{
    (3 +3 k_0r^2)r^2 + a^2
    \big(2\sqrt{(3+3k_0r^2)(3-2r^2)} - (3 - \Lambda r^2)(1+k_0r^2)\big)
    }{a^2 (3 - 2r^2) + (3-\Lambda r^2)(1+k_0r^2)r^2}E,
\end{multline}
and
\begin{equation}
    \label{eq:equaGEO-phiprime}
    \dot\phi
    = \frac{
    (3 - \Lambda r^2)(1+k_0r^2)(L-aE)
    + aE\sqrt{(3+3k_0r^2)(3-2r^2)}
    }
    {
    a^2 (3 - 2r^2) + (3-\Lambda r^2)(1+k_0r^2)r^2
    }.
\end{equation}
Having these last two equations
and setting $g(\dot\gamma,\dot\gamma) = -1$, 
we easily obtain $\dot{r}^2 = V(r)$,
where
\[
V(r) = \frac{V_1(r) + V_2(r) + V_3(r)}{(2r^2 - 3)r^2},
\]
with

\begin{subequations}
    \begin{align}
	V_1(r) &=
	2 a E L
	\left(k_0r^2 (\Lambda  r^2-3)
		+\Lambda r^2 -3 +(3-2r^2) \sqrt{\frac{3 k_0r^2+3}{3-2 r^2}}
	\right),\\
    V_2(r) &=
	(1+k_0r^2)
	\left(
		L^2 (3- \Lambda  r^2)
		-\Lambda  r^4
		-3 \left(E^2-1\right) r^2
	\right),\\
    V_3(r) &= 
    a^2 \left(
	2E^2(2 r^2  - 3)\sqrt{\frac{3 k_0r^2+3}{3-2 r^2}}
	+ E^2(k_0r^2+1) (3 - \Lambda r^2)
	-2r^2
	+3
    \right).
    \end{align}
\end{subequations}

\begin{figure*}[h]
  \centering
    \includegraphics[width=.6\linewidth]{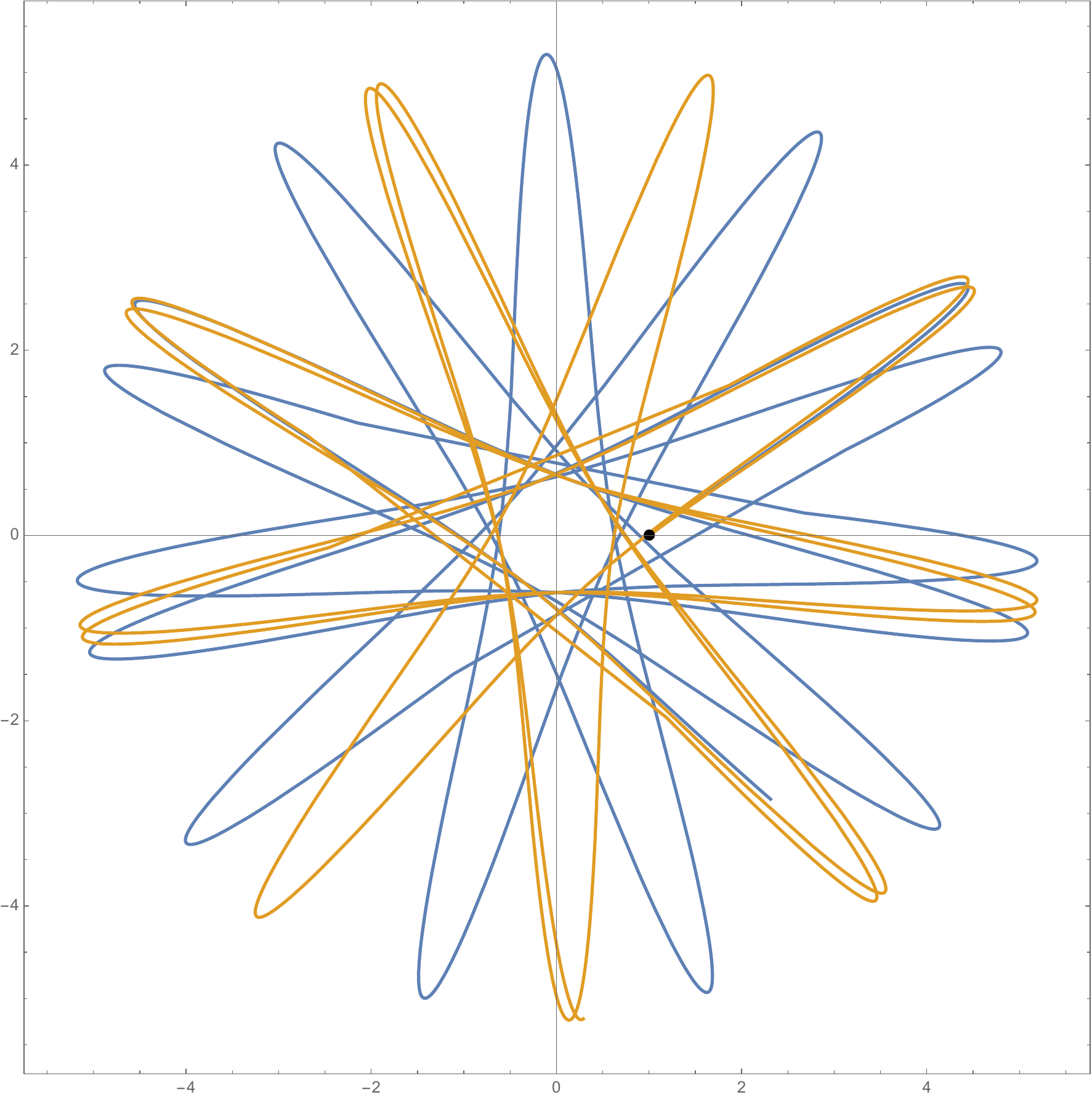}
  \hspace{0mm}\\
  \caption{Same as in Fig. \ref{geo_sph} with respect to the asymmetric metrics obtained via the  GNJ algorithm discussed in Ref.  \cite{aaa,Azreg-Ainou:2014aqa}. The rotation parameter $a$ here is conventionally set to 0.1.}
  \label{geo_ax_eq}
\end{figure*}

Hence, timelike geodesics,
in the equatorial case, not described by circular motions, are solutions of Eq.~\eqref{eq:equaGEO-phiprime}, fulfilling moreover,  

\begin{equation}
\ddot{r} =\frac{1}{2} \frac{\partial V}{\partial r}.    
\end{equation}
A particular case is sketched in Fig. \ref{geo_ax_eq}, where again a comparison is made with another axisymmetric solution -- by analogy with the spherical case, we have chosen the rotating imperfect $\Lambda$--fluid generated by AdS solution after a straightforward application of GNJ algorithm. This metric has once again been described from Ref. \cite{aaa}, and takes the following form,
\begin{multline}\label{eq:AdS-rotating}
g=\left(r^2+\alpha ^2 y^2\right) \cdot\\
\left[
\frac{3}{3 \alpha ^2-\Lambda  r^4+3 r^2}\mathrm dr^2+\frac{2 \alpha  \Lambda  r^4 \left(y^2-1\right) \mathrm dt \mathrm d\varphi }{3 \left(r^2+\alpha ^2 y^2\right)^2}+\frac{ \left(\Lambda  r^4-3 r^2-3 \alpha ^2 y^2\right)}{3 \left(r^2+\alpha ^2 y^2\right)^2}\mathrm dt^2
\right.\\
\left.
+\left(1-y^2\right) \left(1-\frac{\alpha ^2 \left(y^2-1\right) \left(\Lambda  r^4+3 r^2+3 \alpha ^2 y^2\right)}{3 \left(r^2+\alpha ^2 y^2\right)^2}\right)\mathrm d\varphi^2 +\frac{\mathrm dy^2}{1-y^2}
\right]
\end{multline}

\section{Final outlooks}\label{sezione7}

In this paper, we proposed two novel regular spacetimes constructed assuming the presence of a non-zero vacuum energy contribution provided by a (quasi) anti-de Sitter phase. To do so, we started from a spherical solution,  constructed assuming a $00$-component that resembles the anti-de Sitter metric, while inferring a $rr$-component to fulfill the Einstein equations. Thus, requiring pressure isotropy, we found a class of metric quite different from the anti-de Sitter spacetime, albeit with the property of exhibiting a non-zero $\Lambda<0$ term.

From the first spherical spacetime, we assumed the validity of the GNJ algorithm and built up the corresponding cylindrical version, consequently introducing the non-zero vacuum energy term \emph{in addition to} a non-zero quadrupole. For both the metrics, we focused on their kinematic properties and compared with anti-de Sitter solution, both in its classical spherical form and as a rotating \textit{normal} fluid obtained after application of the GNJ algorithm. Specifically, we investigated the geodesic motions for both the two spacetimes under exam. Differences with respect to the standard cases are emphasized throughout the text. Moreover, the property of regularity is emphasized accordingly, showing our two metrics as viable novel regular candidates transporting vacuum energy. 

Further studies will focus on extending our spacetimes through additional external parameters, clarifying the physical role of vacuum energy for each of them. Further \emph{hairy-solutions}, produced from our two hints, will be proposed in our future works.

\section*{Acknowledgements}
D. Corona and R. Giambò thank the partial support of GNAMPA INdAM (Italian National Institute of High Mathematics), Project: CUP-E53C22001930001.
The work of O. Luongo is partially financed by the Ministry of Education and Science of the Republic of Kazakhstan, Grant: IRN AP19680128.

\newpage
\section*{Appendix A: Density and pressure of the axisymmetric case}\label{appA}

The full expressions of $\rho$ and $P=\tfrac{p_1+p_2+p_3}{3}$ are reported below  
\begin{multline}
    \rho(r,y)=
    \frac1{4 \left(r^3+\alpha ^2 r y^2\right)^3 \left(Q(r)r^2+\alpha ^2 y^2\right)^2}\cdot\\
    \left\{r^3 \left(r^2+\alpha ^2 y^2\right)^2 \left[\alpha ^2 r^2 \left(y^2-1\right) \left(r Q'(r)+2 Q(r)\right)^2+8 \alpha ^2 r^2 \left(1-3 y^2\right) Q(r) \right.\right.\\
    \left.
    +4 r Q(r)^2 \left(r^3-\alpha ^2 r y^2 \left(\Lambda  r^2-3\right)\right)+4 \alpha ^4 y^2 \left(y^2-3\right)\right]\\
    +\frac{8}{3} r^3 \left(r^2+\alpha ^2 y^2\right) \left(r^2 Q(r)+\alpha ^2 y^2\right) \cdot\\ 
    \left[\alpha ^2 r^2 Q(r) \left(r y^2 \left(\Lambda  r^2-3\right) Q'(r)+9 y^2-6\right)\right.\\
    \left.
    +Q(r)^2 \left(r^5 \left(\Lambda  r^2-3\right) Q'(r)+3 \alpha ^2 r^2 y^2 \left(\Lambda  r^2-2\right)\right)+3 r^4 Q(r)^3 \left(\Lambda  r^2-2\right)+3 \alpha ^4 y^4\right]\\
    \left.
    -3 r \left(r^2 Q(r)+\alpha ^2 y^2\right)^2 \left(\frac{4}{3} r^6 Q(r)^2 \left(\Lambda  r^2-3\right)-4 \alpha ^2 r^4+4 \alpha ^4 r^2 y^2 \left(y^2-1\right)\right)\right\},\nonumber
\end{multline}

\begin{multline}
    P(r,y)=
    -\frac{1}{36 \left(r^2+y^2 \alpha ^2\right)^3 \left(Q(r) r^2+y^2 \alpha ^2\right)^2}\cdot\\
    \Big\{
    2 r^4 \left(3 \Lambda  r^6+\left(4 y^2 \alpha ^2 \Lambda -1\right) r^4-6 y^2 \alpha ^2 r^2-2 y^4 \alpha ^4\right) Q(r)^4+\\
    8 \left[\left(r^2+y^2 \alpha ^2\right) \left(2 \Lambda  r^4-3 r^2+3 y^2 \alpha ^2\right) Q'(r) r^5+\right.\\
    \left.
    y^2 \alpha ^2 \left(13 \Lambda  r^6+\left(14 y^2 \Lambda  \alpha ^2+3\right) r^4-2 y^2 \alpha ^2 \left(y^2 \alpha ^2 \Lambda -6\right) r^2+18 y^4 \alpha ^4\right) r^2\right] Q(r)^3+\\
    4 Q(r)^2 \left[y^2 \alpha ^2 \Lambda  r^8+3 r^8+30 y^2 \alpha ^2 r^6-6 \alpha ^2 r^6+35 y^4 \alpha ^4 \Lambda  r^6-33 y^4 \alpha ^4 r^4-33 y^2 \alpha ^4 r^4+
    \right.\\
    49 y^6 \alpha ^6 \Lambda  r^4+2 y^2 \alpha ^2 \left(r^2+y^2 \alpha ^2\right)^2 \left(r^2 \Lambda -3\right) Q''(r) r^4+\\
    2 y^2 \alpha ^2 \left(r^2+y^2 \alpha ^2\right) \left(7 \Lambda  r^4-6 r^2+15 y^2 \alpha ^2\right) Q'(r) r^3-\\
     \left.
    66 y^6 \alpha ^6 r^2-27 y^4 \alpha ^6 r^2+12 y^8 \alpha ^8 \Lambda  r^2-6 y^8 \alpha ^8\right] +\\
    4 \alpha ^2 Q(r) \left[-2 y^2 \left(r^2+y^2 \alpha ^2\right)^2 \left(r^2 \Lambda -3\right) Q'(r)^2 r^4+\left(r^2+y^2 \alpha ^2\right)Q'(r) r\cdot
    \right.\\
    \Big(-24 \alpha ^4 y^6+r^2 \alpha ^2 \left(16 \alpha ^2 \Lambda  y^4-63 y^2-15\right) y^2+r^4 \left(26 \alpha ^2 \Lambda  y^4-9 y^2+3\right)\Big)\\ 
    +6 \Big(\left(1-3 y^2\right) r^6+y^2 \left(2 y^2+9\right) \alpha ^2 r^4+2 y^4 \left(y^2+5\right) \alpha ^4 r^2+2 y^6 \alpha ^6\Big)+\\
    \left.
    2 \left(r^3+y^2 \alpha ^2 r\right)^2 \left(r^2 \left(\alpha ^2 \Lambda  y^4+3\right)-3 y^4 \alpha ^2\right) Q''(r)\right] +\\
    \alpha ^2 \Big[12 y^2 \alpha ^2\left(\left(y^2-7\right) r^4+y^2 \left(6 y^2-11\right) \alpha ^2 r^2+2 y^4 \left(y^2-2\right) \alpha ^4+2 \left(r^3+y^2 \alpha ^2 r\right)^2 Q''(r)\right) +\\
    \left(r^3+y^2 \alpha ^2 r\right)^2 \left(r^2 \left(8 \alpha ^2 \Lambda  y^4-9 y^2-39\right)-24 y^4 \alpha ^2\right) Q'(r)^2+\\
    24 \left(4 r y^6 \alpha ^6+11 r^3 y^4 \alpha ^4+7 r^5 y^2 \alpha ^2\right) Q'(r)\Big]
    \Big\},\nonumber
\end{multline}
where $Q(r)$ is the function given by 
\[
Q(r)=\sqrt{\frac{g(r)}{f(r)}}=\sqrt{\frac{1+\kappa  r^2}{1-\frac{2}{3}\Lambda  r^2}}.
\]

\end{document}